\documentclass[12pt]{iopart}
\usepackage{iopams}
\begin{document}

\title{Brownian Motion in a Granular Fluid}

\author{James W. Dufty\dag\ and J. Javier Brey\ddag}

\address{\dag\ Department of Physics, University of Florida, Gainesville, FL 32611, USA}

\address{\ddag\ F\'{\i}sica Te\'{o}rica, Universidad de Sevilla,
Apartado de Correos 1065, E-41080, Sevilla, Spain}

\ead{dufty@phys.ufl.edu, brey@us.es}
\begin{abstract}

The Fokker-Planck equation for a heavy particle in a granular
fluid is derived from the Liouville equation. The host fluid is
assumed to be in its homogeneous cooling state and all
interactions are idealized as smooth, inelastic hard spheres. The
similarities and differences between the Fokker-Planck equation
for elastic and inelastic collisions are discussed in detail.
Although the fluctuation-dissipation relation is violated and the
reference fluid is time dependent, it is shown that diffusion
occurs at long times for a wide class of initial conditions. The
results presented here generalize previous results based on the
Boltzmann-Lorentz equation to higher densities.

\end{abstract}

\submitto{\NJP}


\maketitle

\section{Introduction}

\label{sec1}

In the hundred years since Einstein's remarkable paper on Brownian
motion \cite{Einstein05}, its impact has been felt on a wide range
of problems in Physics, Chemistry, and Mathematics. The subsequent
work of Smoluchowski \cite{Smol06} and Langevin \cite{Langevin08}
extended his physical model to more precise mathematical
structures in the theory of stochastic processes. Attempts to
recover these as idealizations of results from the actual
Newtonian dynamics of the fluid came much later with the
developments of non-equilibrium statistical mechanics
\cite{Lebowitz65}. The detailed description of Brownian motion as
a function of particle size from first principles remains an open
problem \cite{Bocquet00}. A simpler and more controlled problem is
that of a heavy atomic (small size) impurity in an equilibrium
host fluid. Since there is a small parameter, the ratio of the
host fluid particle mass to that of the impurity (and also small
ratio of corresponding mass densities), a systematic analysis
would appear to be straightforward. However, even in this case, a
simple perturbative analysis is found to be limited in accuracy to
its leading asymptotic term and secular terms must be avoided at
higher orders \cite{Cukier69}. The asymptotic analysis for an
impurity in a normal fluid at equilibrium leads to the familiar
Fokker-Planck equation \cite{Lebowitz65,Forster75}. The objective
here is to obtain the extension of that result for an impurity in
a corresponding granular fluid. These results generalize earlier
studies at low density \cite{BDyS99,BRCyD99}.

The idealized prototypical model for a granular fluid is a system
of smooth, inelastic hard spheres. This will be the model
considered here. The inelasticity of collisions implies a
continual loss of energy on each collision. Consequently, the
usual equilibrium state for an isolated molecular system is
replaced by a homogeneous ``cooling'' state (HCS), which serves as
the environment in which the motion of a heavy impurity is
considered. Although this reference HCS is inherently time
dependent, an equivalent stationary representation can be given
using appropriate velocity scaling. The time dependence of the
probability density for the impurity particle in its phase space
of position and velocity, can then be analyzed by methods similar
to those for normal fluids. For example, the general case of
diffusion of an impurity particle of arbitrary mass in a fluid
undergoing HCS has been given recently in \cite{Dufty02}. In this
work, a Zwanzig-Mori projection operator method \cite{Zw01} is
used (see \ref{ap1}) to obtain a formally exact kinetic equation
vaild for this general case. The Fokker-Planck equation then
follows from this kinetic equation to leading order in the ratio
of a fluid particle mass to the impurity mass. The analysis is
complicated by the singular nature of the hard sphere interaction,
but this mass ratio expansion has been discussed in detail for
hard, elastic collisions \cite {Bocquet94}. An interesting new
complication for inelastic collisions is a restriction on the
cooling rate of the host fluid relative to the collision frequency
of the impurity. If this is too large, the expansion is no longer
valid. This point has already been noted and discussed elsewhere
\cite{BDyS99,SyD01}.

The rest of the presentation focuses on the similarities and
differences between impurity motion in normal and granular fluids.
First it is shown that there is an exact mapping of the granular
Fokker-Planck equation onto that for elastic collisions, in
appropriate dimensionless variables. Thus the physical mechanisms
of fast velocity relaxation to a Maxwellian for homogeneous
initial states, and fast relaxation to a diffusive equation for
inhomogeneous states occurs for granular gases as well. In
particular, the transition to a hydrodynamic stage (diffusion)
takes place for all degrees of inelasticity. However, in terms of
physical variables there are significant differences. One of these
is the violation of equipartition, since the kinetic temperature
of the impurity is different from that of the fluid in the
stationary state. Also, the usual fluctuation-dissipation relation
between the drift and diffusion coefficients no longer holds, as
expected for a non-equilibrium state of the host fluid. One
consequence is that the mobility coefficient, measuring the
response to an external force (for instance, electromagnetic or
gravitational), is not simply related to the diffusion coefficient
by the usual Einstein formula.

The Liouville equation for a granular fluid \cite{BDyS97,NyE01} is
introduced in the next Section. The special HCS for an isolated
system is described and its time dependence is removed through the
introduction of dimensionless variables. It is shown that the HCS
is a stationary state for the Liouville equation in this
representation. This stationary representation is used to describe
the motion of an impurity in the fluid. A formally exact kinetic
equation is derived for the impurity particle in Sec \ref{s3}, and
the condition for its stationary solution is described. Next, in
Section \ref{s4} the form of this kinetic equation is simplified
to that of the usual Fokker-Planck equation \cite{McL89} in the
asymptotic limit of small ratio of fluid mass to impurity mass.
The details of the reduction are given in the Appendices. Some of
the most important similarities and differences between impurity
motion in normal and granular fluids are listed in Section
\ref{s5}, and summary remarks are given in the last Section of the
paper.

\section{The Homogeneous Cooling State and its Stationary Representation}
\label{sec2}

The basis for the analysis to be carried out in this paper is the
Liouville equation for a fluid of $N$ smooth, inelastic hard
spheres ($d=3$) or disks ($d=2$) of mass $m$ and diameter
$\sigma$, and one impurity of mass $m_{0}$ and diameter
$\sigma_{0}$. The formal structure of this equation has been
discussed in detail elsewhere in the general context of impurity
diffusion \cite{Dufty02}, and only an overview will be given here.
The position and velocity coordinates of the $N$ equal particles
will be denoted by $\{ {\bi q}_{i},{\bi v}_{i}; i=1, \cdots, N
\}$, while those of the impurity particle by ${\bi q}_{0},{\bi
v}_{0}$. The expectation value for some observable $A(\Gamma )$
defined over the $N+1$ particle phase space $\Gamma \equiv \{ {\bi
q}_{0},{\bi q}_{1}, \cdots, {\bi q}_{N},{\bi v}_{0},{\bi
v}_{1},\cdots,{\bi v}_{N} \}$, is given by
\begin{equation}
\langle A;t\rangle =\int d\Gamma\, \rho (\Gamma ,t)A(\Gamma ),
\label{2.1}
\end{equation}
where $\rho(\Gamma,t)$ is the probability density for the state of
the system at time $t$ and the dynamics is determined from the
Liouville equation
\begin{equation}
\left( \partial _{t}+\overline{L}\right) \rho (\Gamma ,t)=0.
\label{2.2}
\end{equation}
The generator for the inelastic hard particle dynamics is
\begin{eqnarray}
\overline{L} &=& \overline{L}_{f}+\mathbf{v}_{0} \cdot
\mathbf{\nabla}_{0}-\sum_{i=1}^{N}\overline{T}_{-}(0,i), \nonumber
\\
\overline{L} _{f} &=& \sum_{i=1}^{N} \mathbf{v}_{i}\cdot
\mathbf{\nabla}_{i}-\frac{1}{2}\sum_{i=1}^{N} \sum_{j \neq i}^{N}
\overline{T}_{-}(i,j). \label{2.3}
\end{eqnarray}
Here, $\overline{L}_{f}$ is the generator for the dynamics of the
fluid particles alone and the subscripts or labels $0$ refer to
the impurity particle. The second term in the expression of
$\overline{L}$ generates the free streaming of the impurity. The
operators $\overline{T}_{-}(i,j)$ and $\overline{T}_{-}(0,i)$
describe binary scattering between fluid particles and between the
impurity and fluid particles, respectively,
\begin{equation}
\fl \overline{T}_{-}(i,j)=\sigma ^{d-1}\int d\Omega \,\Theta
(\mathbf{g}_{ij}\cdot \widehat{\mbox{\boldmath
$\sigma$}})|\mathbf{g}_{ij}\cdot \widehat{ \mbox{\boldmath
$\sigma$}}|[\alpha ^{-2}\delta (\mathbf{q}_{ij}- \mbox{\boldmath
$\sigma$})b_{ij}^{-1}-\delta (\mathbf{q}_{ij}+ \mbox{\boldmath
$\sigma$})], \label{2.4}
\end{equation}
\begin{equation}
\fl \overline{T}_{-}(0,i)=\overline{\sigma }^{d-1}\int d\Omega
\,\Theta (\mathbf{g} _{0i}\cdot \widehat{\mbox{\boldmath
$\sigma$}})|\mathbf{g}_{0i}\cdot \widehat{\mbox{\boldmath
$\sigma$}}|[\alpha _{0}^{-2}\delta (\mathbf{q}_{0i}-
\overline{\mbox{\boldmath $\sigma$}})b_{0i}^{-1}-\delta
(\mathbf{q}_{0i}+ \overline{\mbox{\boldmath $\sigma$}})],
\label{2.5}
\end{equation}
where $d\Omega $ is the solid angle element for the unit vector $
\widehat{\mbox{\boldmath $\sigma$}}$, $\mbox{\boldmath
$\sigma$}=\sigma \widehat{\mbox{\boldmath $\sigma$}}$, $\Theta $
is the Heaviside step function, and $\mathbf{g} _{ij}={\bi
v}_{i}-{\bi v}_{j}$ and $\mathbf{q}_{ij}=\mathbf{q}_{i}-
\mathbf{q}_{j}$ are the relative velocities and positions,
respectively. Moreover, $\overline{ \mbox{\boldmath
$\sigma$}}=\overline{\sigma }\widehat{\mbox{\boldmath $\sigma$}}$
with $\overline{\sigma }=\left( \sigma +\sigma _{0}\right) /2$ .
The operators $b_{ij}^{-1}$ and $b_{0i}^{-1}$ determine the
pre-collision velocities in a restituting collision. They are
defined by
\begin{eqnarray}
b_{ij}^{-1}{\bi v}_{i} & = & {\bi v}_{i}-\frac{1+\alpha}{2 \alpha}
( \mathbf{g}_{ij} \cdot \widehat{\bsigma})
\widehat\bsigma, \nonumber \\
b_{ij}^{-1} {\bi v}_{j}  & = & {\bi v}_{j}+\frac{1+\alpha}{2
\alpha} ( \mathbf{g}_{ij} \cdot \widehat{\bsigma})
\widehat\bsigma, \label{2.6}
\end{eqnarray}
and
\begin{eqnarray}
b_{0i}^{-1} {\bi v}_{0} & =& {\bi v}_{0}-
\frac{(1+\alpha_{0})\Delta}{\alpha_{0}}\, ({\bi g}_{0i} \cdot
\widehat\bsigma) \widehat\bsigma, \nonumber \\
b_{0i}^{-1} {\bi v}_{i}& = & {\bi v}_{i}+ \frac{(1+\alpha_{0})(1-
\Delta)}{\alpha_{0}}\, ({\bi g}_{0i} \cdot \widehat\bsigma)
\widehat\bsigma. \label{2.7}
\end{eqnarray}
We have introduced $\Delta =m/\left( m_{0}+m\right)$, that is the
small parameter of the system in the Brownian limit to be
discussed in the following. In addition, $\alpha$ and $\alpha_{0}$
are the coefficients of normal restitution for fluid-fluid and
impurity-fluid collisions, respectively. These coefficients
measure  the degree of inelasticity of collisions and take values
in the interval $0< \alpha, \alpha_{0} \leq 1$.

Average kinetic (granular) temperatures for the fluid, $T(t)$, and
the impurity particle, $T_{0}(t)$, are defined by
\begin{equation}
T(t)\equiv \frac{1}{Nd} \sum_{i=1}^{N} m \langle {\bi v}
_{i}^{2};t \rangle \equiv \frac{1}{2}m \widetilde{v}^{2}(t),
\label{2.8}
\end{equation}
\begin{equation}
T_{0}(t)\equiv \frac{1}{d} m_{0} \langle {\bi v}_{0}^{2};t\rangle
\equiv \frac{1}{2}m_{0} \widetilde{v}_{0}^{2}(t). \label{2.9}
\end{equation}
The average velocities $\widetilde{v}(t)$ and
$\widetilde{v}_{0}(t)$ introduced above will be called thermal
velocities of the fluid and the impurity, respectively. Since all
the collisions are inelastic and the effect of the impurity on the
fluid is negligible in the limit of large $N$, $T(t)$ decreases
monotonically in time for an isolated system with $\alpha <1$.
More precisely,
\begin{equation}
\partial _{t}T(t)=-\zeta (t)T(t),
\label{2.10}
\end{equation}
where $\zeta (t)$ is the ``cooling'' rate of the fluid due to
inelastic collisions,
\begin{equation}
\zeta \left( t\right) =- \frac{2N}{\widetilde{v}^{2}(t)d}\int
d\Gamma\,  {\bi v}_{1}^{2}\, \overline{T}_{-}(1,2)\rho (\Gamma ,t)
\geq 0. \label{2.11}
\end{equation}
Similarly, the time evolution of $T_{0}$ is given by
\begin{equation}
\partial _{t}T_{0}(t)=-\zeta _{0}(t)T_{0}(t) \label{2.12}
\end{equation}
with
\begin{equation}
\zeta _{0}(t) =-\frac{2N}{\widetilde{v}_{0}^{2}(t) d}\int
d\Gamma\, {\bi v}_{0}^{2}\,\overline{T} _{-}(0,1)\rho (\Gamma ,t),
\label{2.13}
\end{equation}
but it can not be concluded that the impurity temperature is
monotonically decreasing, in general. Its behavior depends in
detail on its initial value relative to that of the fluid
particles.

Since the fluid is cooling, there is no stationary solution to the
Liouville equation for an isolated system. Instead, it is
postulated that there is a solution whose time dependence occurs
entirely through $T(t)$ and $T_{0}(t)$, having the scaling
property
\begin{equation}
\rho _{hcs}(\Gamma ,t)=\left[ \ell \widetilde{v}_{0,hcs}(t)\right]
^{-d}\left[ \ell \widetilde{v}_{hcs}(t)\right] ^{-Nd} \rho
_{hcs}^{\ast }\left( \{\mathbf{q}_{ij}^{\ast }, {\bi v}_{i}^{\ast
},\mathbf{q}_{i0}^{\ast },{\bi v}_{0}^{\ast }\}\right),
\label{2.14}
\end{equation}
with the dimensionless variables defined by
\begin{equation}
\mathbf{q}_{i}^{\ast }=\mathbf{q}_{i}/\ell,  \quad
\mathbf{q}_{0}^{\ast }=\mathbf{q}_{0}/\ell,  \quad {\bi
v}_{i}^{\ast }= {\bi v}_{i}/\widetilde{v}_{hcs}(t),  \quad {\bi
v}_{0}^{\ast }={\bi v} _{0}/\widetilde{v}_{0,hcs}(t). \label{2.15}
\end{equation}
The coordinates have been scaled relative to $\ell \equiv (n\sigma
^{d-1})^{-1}$, where $n$ is the number density of particles, so
that $\ell$ is proportional to the mean free path. The subscripts
$hcs$ on the thermal velocities denote their values calculated for
this particular solution. The dimensionless distribution function
$\rho _{hcs}^{\ast }$ is invariant under space translations and
therefore $\rho_{hcs}(\Gamma,t)$ represents a spatially
homogeneous state of the fluid plus the impurity. The scaling form
in equation (\ref{2.14}) implies certain constraints on the
cooling rates. Let us introduce reduced cooling rates by
\begin{equation}
\zeta _{hcs}^{\ast } \equiv \frac{\ell \zeta _{hcs}(t)
}{\widetilde{v}_{hcs}(t)}, \quad \quad \quad \zeta _{0,hcs}^{\ast
}\left[ \gamma (t)\right] \equiv \frac{\ell \zeta
_{0,hcs}(t)}{\widetilde{v}_{0,hcs}(t)}\, . \label{2.16}
\end{equation}
From Eqs.\ (\ref{2.11}) and (\ref{2.14}) it is easily seen that
the reduced cooling rate of the fluid, $\zeta^{*}_{hcs}$, is time
independent. Moreover, the cooling rate for the impurity in this
state, $\zeta_{0,hcs}^{*}$, depends on time only through the ratio
of temperatures
\begin{equation}
\gamma ( t)  \equiv
\frac{\widetilde{v}_{0,hcs}(t)}{\widetilde{v}_{hcs}(t)}=\sqrt{\frac{
T_{0}\left( t\right) m}{T(t)m_{0}}}\, .  \label{2.17}
\end{equation}
Similarly, it is found that the Liouville equation for
$\rho_{hcs}^{*}$ in these dimensionless variables depends
explicitly on time only through $ \gamma (t)$. Since
$\rho_{hcs}^{\ast }$ is time independent, solutions exist only if
$\gamma (t)$ is a constant. The cooling rate equations (\ref
{2.10}) and (\ref{2.12}) then give immediately the requirement
\begin{equation}
\zeta _{hcs}^{\ast }=\gamma \zeta _{0,hcs}^{\ast }\left( \gamma
\right) . \label{2.18}
\end{equation}

In the particular state we are considering, the fluid particles
and the impurity particle cool at the same rate, i.e.
$\zeta_{hcs}(t)=\zeta_{0,hcs}(t)$. This should be viewed as a
condition for the existence itself of this state as it fixes one
of the two temperatures in terms of the other. However, this does
not imply of course that the two temperatures are the same. In
fact, it is found that they must be different except in the case
of mechanically identical particles, as discussed below. With
these results, the equation for $\rho _{hcs}^{\ast }$ becomes
independent of time and has the form:
\begin{equation}
\frac{1}{2}\zeta _{hcs}^{\ast }\sum_{i=0}^{N}
\frac{\partial}{\partial {\bi v}^{*}_{i}}\cdot ({\bi v}_{i}^{\ast
}\rho _{hcs}^{\ast })+\overline{L}^{\ast }\left( \gamma \right)
\rho _{hcs}^{\ast }=0,  \label{2.19}
\end{equation}
where $\overline{L}^{\ast }$ is the Liouville operator in
dimensionless form (see below). The self-consistent solution to
the coupled set of equations formed by equation (\ref{2.19}) and
the cooling equations determines the \emph{homogeneous cooling
state} (HCS). It is the analogue of the Gibbs state for molecular
systems and reduces to it for $\alpha = \alpha_{0}=1$.

In dimensionless form, as defined by $\rho^{*}_{hcs}$, the HCS is
time independent. This suggests a transformation of the Liouville
equation to dimensionless variables such that the HCS is a
stationary solution in that representation. This can be formally
carried out in the following way. We are going to scale the
positions and velocities as given by Eqs.\ (\ref{2.15}), where now
$\widetilde{v}_{hcs}(t)$ is defined by
$\widetilde{v}_{hcs}(t)=(2T_{hcs}/m)^{1/2}$ and $T_{hcs}(t)$ is
the solution of the equation
\begin{equation}
\partial_{t} T_{hcs}(t)=-\zeta_{hcs}(t) T_{hcs}(t),
\label{2.20}
\end{equation}
with the initial condition $T_{hcs}(t_{0})=T(t_{0})$, $t_{0}$
being some arbitrary time. Of course, since in the general case
the system is not in the HCS, it is $T(t) \neq T_{hcs}(t)$.
Moreover, in the scaling $\widetilde{v}_{0,hcs}(t)$ is chosen as
given by $\widetilde{v}_{0,hcs}=\gamma \widetilde{v}_{hcs}(t)$,
where $\gamma$ is the time-independent parameter identified by
equation (\ref{2.18}). The appropriate dimensionless time scale
$s$ is proportional to the average accumulated number of number of
collisions for the fluid particles in the reference HCS,
\begin{equation}
s(t,t_{0})=\int_{t_{0}}^{t}dt^{\prime }\, \frac{v_{hcs}(t^{\prime
})}{\ell }. \label{2.21}
\end{equation}
For a general state whose distribution function is $\rho
(\Gamma,t)$, the dimensionless distribution
$\rho^{*}(\Gamma^{*},s)$ is defined by
\begin{equation}
\rho (\Gamma ,t)=\left[ \ell \widetilde{v}_{0,hcs}(t)\right]
^{-d}\left[ \ell \widetilde{v}_{hcs}(t) \right] ^{-Nd}\rho
^{*}(\Gamma^{*},s) , \label{2.22}
\end{equation}
where $\Gamma^{*} \equiv \{ {\bi q}^{*}_{0}, \cdots , {\bi
q}^{*}_{N},{\bi v}^{*}_{0}, \cdots, {\bi v}^{*}_{N}\}$.
Substitution of this into equation (\ref{2.2}) gives the
dimensionless Liouville equation
\begin{equation}
( \partial _{s}+\overline{\mathcal{L}}^{*}) \rho ^{\ast }(\Gamma
^{\ast },s)=0,  \label{2.23}
\end{equation}
with the operator $\overline{\mathcal{L}^{*}}$ defined by
\begin{eqnarray}
\fl \overline{\mathcal{L}}^{\ast }\Phi (\Gamma^{*}) =\left[ \gamma
{\bi v}_{0}^{\ast }\cdot \mathbf{\nabla}^{*}_{0}- \gamma
\sum_{i=1}^{N}\overline{T}^{\ast }_{-}(\gamma
;0,i)+\frac{1}{2}\zeta _{hcs}^{*} \frac{\partial}{\partial {\bi
v}_{0}^{*}} \cdot {\bi v}_{0}^{*} \right. \nonumber \\
\left. +\overline{L}_{f}^{*}+\frac{1}{2}\zeta
_{hcs}^{*}\sum_{i=1}^{N} \frac{\partial}{\partial {\bi v}_{i}^{*}}
\cdot {\bi v}_{i}^{*} \right] \Phi (\Gamma^{*}), \label{2.24}
\end{eqnarray}
for arbitrary $\Phi(\Gamma^{*})$. The detailed form of the
dimensionless binary collision operators $\overline{T}^{\ast
}(\gamma ;0,i)$ is given in \ref{ap1}. The first three terms on
the right side of equation (\ref{2.24}) describe the dynamics of
the impurity particle. Since the derivative with respect to $s$ in
(\ref {2.23}) is taken at constant ${\bi v}_{0}^{\ast }$ and ${\bi
v}_{i}^{\ast }$, some effects of cooling become explicit in the
terms proportional to the cooling rate. The last two terms on the
right-hand-side of (\ref{2.24}) generate the dynamics of the fluid
without the impurity, again with explicit effects of cooling. As a
consequence of these cooling terms, it is seen that the
consistency condition for a stationary solution is the same as
equation (\ref{2.19}) for the HCS solution,
\begin{equation}
\overline{\mathcal{L}}^{\ast }\rho _{hcs}^{\ast }=0,  \label{2.25}
\end{equation}
since in equation (\ref{2.19}) it is:
\begin{equation}
\overline{L}^{*}(\gamma)=L_{f}^{*}+\gamma {\bi v}^{*}_{0} \cdot
\nabla_{0} -\gamma \sum_{i=1}^{N}
\overline{T}_{-}^{*}(\gamma;0,i). \label{2.26}
\end{equation}
Therefore, the dimensionless form of the Liouville equation,
equation (\ref{2.24}), will be referred to as the stationary
representation since it supports a stationary state and that state
is the HCS.

\section{Kinetic Equation for the Impurity Particle}
\label{s3}

Consider now the case of the fluid and the impurity in the HCS. At
some instant (taken to be $t=t_{0}=0$) the impurity is observed to
be at position $\mathbf{R}_{0}$ with velocity ${\bi V}_{0}$. This
initial state can be represented as
\begin{equation}
\rho \left( \Gamma ,0 \right) =\frac{\rho _{hcs} \left( \Gamma
\right) \delta \left( \mathbf{R}_{0}-\mathbf{q}_{0} \right) \delta
\left( {\bi V}_{0}- {\bi v}_{0} \right)}{\int d\Gamma \rho
_{hcs}\left( \Gamma \right) \delta \left(
\mathbf{R}_{0}-\mathbf{q}_{0} \right) \delta \left( {\bi V} _{0}-
{\bi v}_{0}\right)}\, . \label{3.1}
\end{equation}
This is equivalent to
\begin{equation}
\rho^{*} \left( \Gamma ^{*},0\right) =\frac{\rho_{hcs}^{*}\left(
\Gamma^{*} \right) \delta \left(
\mathbf{R}_{0}^{*}-\mathbf{q}_{0}^{*}\right) \delta \left( {\bi
V}_{0}^{*} -{\bi v}^{*}_{0}\right) }{\int d\Gamma^{*} \rho
_{hcs}^{*} \left( \Gamma \right) \delta \left( \mathbf{R}_{0}^{*}-
\mathbf{q}_{0}^{*} \right) \delta \left( {\bi V}_{0}^{*}- {\bi
v}_{0}^{*} \right) }\ \equiv \rho^{*}_{hcs}(\Gamma^{*};X_{0}^{*}).
\label{3.2}
\end{equation}
The second equality introduces a convenient notation for the state
in which the impurity variables $X \equiv \{ {\bi R}, {\bi V} \}$
are sharply defined. \emph{In the remainder of this discussion,
the stationary representation in the dimensionless units will be
used, although for simplicity of notation the asterisks will be
omitted}. The probability $F({\bi R},{\bi V},s)$ for the impurity
to have position and velocity ${\bi R},{\bi V}$, at ``time'' $s$
is then
\begin{equation}
\fl F(X,s)=\int d\Gamma\, \delta (X-x_{0}) \rho (\Gamma,t) = \int
d\Gamma \delta \left( X-x_{0}\right)
e^{-\overline{\mathcal{L}}s}\rho _{hcs}\left( \Gamma
;X_{0}\right), \label{3.3}
\end{equation}
where, trivially, $x_{0} \equiv \{ {\bi q}_{0},{\bi v}_{0} \}$. A
formally exact kinetic equation for this probability is obtained
in \ref{ap1}. It has the form
\begin{eqnarray}
\fl \left[ \frac{\partial}{\partial s}+\gamma {\bi V}\cdot
\frac{\partial}{\partial {\bi R}}+ \gamma \Lambda
+\frac{\zeta_{hcs}}{2} \frac{\partial}{\partial {\bi V}} \cdot
{\bi V} \right] F(X,s) \nonumber \\
+\int_{0}^{s}ds^{\prime }\int dX^{\prime }M(X,X^{\prime
},s-s^{\prime })F(X^{\prime },s^{\prime })=0. \label{3.4}
\end{eqnarray}
The linear operator\ $\Lambda $, defined in equation (\ref{a.16})
in \ref{ap1}, describes the exact short time effect of collisions,
including the initial correlations among particles in the HCS. If
velocity correlations are neglected, $\Lambda $ reduces to the
Enskog-Lorentz collision operator. The form of the operator
$M(X,X^{\prime },s)$ is given by equation (\ref{a.27}). It
describes dynamically correlated collisions of the impurity with
two or more particles that develop in time. These contributions
are vanishingly small at short times and generally negligible at
low densities. However, at very high densities, they describe the
dominant effects of ``caging'' as the impurity particle becomes
more localized due to collisions.

The stationary HCS, $F_{hcs}({\bi V})$, is determined from
equation (\ref{3.4}) by
\begin{equation}
\frac{\zeta _{hcs}}{2} \frac{\partial}{\partial {\bi V}} \cdot
\left[ {\bi V} F_{hcs}({\bi V}) \right] + \gamma \Lambda
F_{hcs}({\bi V})=0. \label{3.5}
\end{equation} Use has been made of the property given by equation (\ref{a.18}),
i.e.
\begin{equation}
\int dX^{\prime }M(X,X^{\prime },s)F_{hcs}({\bi V}^{\prime })=0.
\label{3.6}
\end{equation}
Thus, the HCS is a stationary point of $M$ but not of $\Lambda $.
Further comment on this is given in the next section. Also, the
particularization for the HCS of the expression for the cooling
rate given by equation (\ref{2.13}) can be expressed in terms of
$F_{hcs}$ as
\begin{equation}
\zeta _{0,hcs}=\frac{2 W}{d}\int d{\bi v}_{0}\, {\bi v}_{0}^{2}
\Lambda F_{hcs}({\bi v}_{0}), \label{3.7}
\end{equation}
$W$ being the volume of the system. The results of this Section
are still exact and serve as a suitable starting point for the
small mass ratio expansions to be discussed in the next Section.

\section{Fokker-Planck Limit}
\label{s4}

In this Section, it will be shown that the kinetic equation
(\ref{3.4}) reduces to a Fokker-Planck equation for asymptotically
small $m/m_{0}$. The analysis proceeds in the following way.
First, it is shown in \ref{ap2} that the operator $\Lambda $
becomes in the limit $m/m_{0}\rightarrow 0$
\begin{equation}
\gamma \Lambda F(X,s)=- \Gamma_{0} \frac{\partial}{\partial {\bi
V}} \cdot \left( {\bi V}+ \frac{G}{2} \frac{\partial}{\partial
{\bi V}} \right) F(X,s), \label{4.1}
\end{equation}
where $\Gamma_{0} $ and $G$ are collision integrals, independent
of time and velocity, given by Eqs.\ (\ref{b.17}) and
(\ref{b.20}), respectively,
\begin{equation}
\Gamma_{0}=\frac{(1+\alpha_{0})\Delta}{2} \left(
\frac{\overline{\sigma}}{\sigma} \right)^{d-1} C_{1}, \label{4.2}
\end{equation}
\begin{equation}
G=\frac{\gamma^{-2} (1+\alpha_{0}) \Delta}{2}\, C_{2}, \label{4.3}
\end{equation}
$C_{1}$ and $C_{2}$ being real, positive numbers, of the order of
unity. Note that in the limit we are considering, $\gamma^{-2}
\Delta$ is a constant, independent of the mass ratio. For elastic
collisions, $G\rightarrow 1$ and $\Lambda $ becomes the usual
Fokker-Planck operator in dimensionless form. The fact that $G\neq
1$ here can be considered as a violation of a
fluctuation-dissipation relation.

\subsection{Impurity HCS}

Let us analyze the consequences of this for the impurity HCS. Use
of equation (\ref {4.1}) into  equation (\ref{3.5}) gives
immediately
\begin{equation}
\frac{\partial}{\partial {\bi V}} \cdot \left[ \left( 1-\epsilon
\right) {\bi V}+\frac{1}{2}G \frac{\partial}{\partial {\bi V}}
\right] F_{hcs}=0, \label{4.4}
\end{equation}
where we have introduced the parameter
\begin{equation}
\epsilon \equiv \frac{\zeta _{hcs}}{2\Gamma _{0}}. \label{4.5}
\end{equation}
It is seen that the solution to this equation is a Gaussian.
Normalization and the definition of the impurity temperature
require that
\begin{equation}
F_{hcs}({\bi V})=\pi ^{-d/2}e^{-V^{2}}.  \label{4.6}
\end{equation}
Then equation (\ref{4.4}) gives the relation
\begin{equation}
G=1-\epsilon.  \label{4.7}
\end{equation}
This allows determination of the impurity temperature in the HCS
in terms of the fluid temperature. The latter is determined from
equation (\ref{2.10}) using for $\zeta_{hcs}$ the expression
identified bellow, and has a universal value, depending only on
$\alpha $ and  being independent of the initial conditions at long
times. The impurity temperature in the HCS is then determined from
equation (\ref{4.7}) with $G$ given by equation (\ref{4.3}),
\begin{equation}
T_{0}(t)=\frac{\left( 1+\alpha _{0}\right)C_{2} }{2\left(
1-\epsilon \right) } T(t), \label{4.8}
\end{equation}
following that $T_{0}(t)\neq T(t)$.

A further consequence of the above relationship is the condition
$\epsilon <1$, i.e.
\begin{equation}
\frac{\zeta _{hcs}}{2\Gamma _{0}}<1.  \label{4.9}
\end{equation}
Thus the small mass ratio limit for granular fluids entails, in
addition, a restriction on the inelasticity of the fluid-fluid
particle collisions. This follows because $\zeta _{hcs}\propto
1-\alpha $ and $\Gamma_{0} \propto \Delta $, so the fluid
inelasticity must decrease as the mass ratio goes to zero.
Asymptotically, for small $1-\alpha $ it is found that (see
\ref{ap3}) \begin{equation} \zeta _{hcs} \sim \frac{4
\pi^{(d-1)/2}}{\sqrt{2} \Gamma(d/2) d} (1-\alpha) \chi(\sigma),
\label{4.10}
\end{equation}
where $\chi(\sigma)$ is the fluid-fluid pair correlation function
for particles at contact and $\Gamma(y)$ is the Euler Gamma
function. Then the condition (\ref{4.9}) becomes
\begin{equation}
\frac{4 \pi^{(d-1)/2} \chi (\sigma)
 (1-\alpha)}{ \sqrt{2} \Gamma (d/2) C_{1} d (1+\alpha_{0}) \Delta}\,
\left( \frac{\sigma}{\overline{\sigma}} \right)^{d-1} < 1.
 \label{4.11}
\end{equation}
In practice, for any choice of $\Delta <<1$ the value of $\alpha$
must verify $1-\alpha <<1 $, moreover being consistent with
equation (\ref{4.11}). Otherwise, the derivation of the
Fokker-Planck limit for $\Lambda $ does not hold. The correct
limit for $ \epsilon >1$ is described in reference \cite{SyD01}.
Equation (\ref{4.11}) can be made more explicit in the low density
limit of the fluid. Then, the correlation function $g_{0}^{(2)}$
appearing in the expression of $C_{1}$, equation (\ref{b.18}), can
be set equal to unity, and $g_{1}^{(2)}$ can be neglected. In this
way, it is obtained
\begin{equation}
\frac{1-\alpha}{\sqrt{2} (1+\alpha_{0}) \Delta} \left(
\frac{\sigma}{\overline{\sigma}} \right)^{d-1} < 1, \label{4.12}
\end{equation}
that agrees with the result derived from the Lorentz-Boltzmann
equation reported in reference \cite{BDyS99}.

In summary, the stationary state condition fixes the form of
$\Lambda $ as a Fokker-Planck operator with friction constant
$\Gamma _{0}$ and diffusion coefficient $(1-\epsilon ),$
\begin{equation}
\Lambda F=- \gamma^{-1} \Gamma _{0} \frac{\partial}{\partial {\bi
V}} \cdot \left[ \mathbf{V}+ \frac{1}{2}(1-\epsilon )
\frac{\partial}{\partial {\bi V}} \right] F, \label{4.13}
\end{equation}
where $\epsilon$ has been defined in equation (\ref{4.3}).
Moreover, the stationary state equation (\ref{4.4}) simplifies to
\begin{equation}
\left( 1-\epsilon \right) \frac{\partial}{\partial {\bi V}} \cdot
\left( \mathbf{V }+\frac{1}{2}\frac{\partial}{\partial {\bi V}}
\right) F_{hcs}({\bi V})=0. \label{4.14}
\end{equation}

\subsection{Fokker-Planck Equation}

With the form for $\Lambda $ given in equation (\ref{4.13}), the
kinetic equation (\ref{3.4})becomes
\begin{eqnarray}
\fl \left( \partial _{s}+\gamma \mathbf{V}\cdot \mathbf{\nabla
}_{\mathbf{R} }\right) F(X,s) =\left( 1-\epsilon \right) \Gamma
_{0} \frac{\partial}{\partial {\bi V}} \cdot \left(
\mathbf{V}+\frac{1}{2} \frac{\partial}{\partial {\bi V}}
\right) F(X,s)  \nonumber \\
+\int_{0}^{s}ds^{\prime }\int dX^{\prime }M(X,X^{\prime },s-
s^{\prime })F(X^{\prime },s^{\prime }).  \label{4.15}
\end{eqnarray}
The first term on the right side is now a Fokker-Planck operator
of the usual form, but the ``friction constant'' $\Gamma _{0}$
obtained from $\Lambda $, has been renormalized by $\left(
1-\epsilon \right) $, due to the additional operator representing
cooling.

The values of $\gamma ,$ $\epsilon $, and $\Gamma_{0} $ are now
known and it remains only to determine the asymptotic form of the
last term on the right hand side of equation (\ref{4.15}). There
are two steps in that process. First, since $M(X,X^{\prime
},s^{\prime })$ involves two binary collision operators it turns
out to be proportional to $\left( \Delta \gamma ^{-1}\right)
^{2}$. By exploiting this, in \ref{ap2} it is found that
\begin{equation}
\fl \int dX^{\prime} M(X,X^{\prime},s-s^{\prime})
F(X^{\prime},s^{\prime})= \frac{\partial}{\partial {\bi V}} \cdot
\left(  {\bi V}+\frac{1}{2}\, \frac{\partial}{\partial {\bi V}}
\right) \mathcal{G}_{c}(s-s^{\prime}) F(X,s^{\prime}),
\label{4.16}
\end{equation}
where $\mathcal{G}_{c}(s)$ is the correlation function
\begin{equation}
\mathcal{G}_{c}(s)=\frac{2}{d} \left\langle {\bi F}_{+}(s)\cdot
{\bi F} _{-}\right\rangle _{f}. \label{4.17}
\end{equation}
The brackets denote an equilibrium ensemble average for the fluid
in the presence of an infinitely heavy impurity at rest. The phase
functions ${\bi F}_{\pm }$ are the total rate of momentum transfer
by the fluid particles to the impurity,
\begin{equation}
{\bi F}_{\pm }= \gamma^{-1} \left( 1+\alpha _{0}^{\pm 1}\right)
\Delta \overline{\sigma}^{d-1} \sum_{i=1}^{N}\int d\Omega \,\Theta
( \mp {\bi v}_{i}\cdot \widehat{\mbox{\boldmath $\sigma$}} )\delta
(\mathbf{q}_{0i}+ \overline{\mbox{\boldmath $\sigma$}})( {\bi
v}_{i}\cdot \widehat\bsigma)^{2}\widehat\bsigma. \label{4.18}
\end{equation}

For continuous potentials they would be simply the total force on
the impurity, but are determined here for hard spheres by the
binary collision operators. The final step in the reduction of
term involving $M(X,X^{\prime },s-s^{\prime })$ is to recognize
that the time scale for the impurity as described through $F(X,s)$
is slower by a factor $\gamma $ than the decay time for the fluid
correlation function $\mathcal{G}_{c}(s)$. To leading order in the
mass ratio this gives
\begin{equation}
\int_{0}^{s}ds^{\prime }\mathcal{G}_{c}(s- s^{\prime })F(X^{\prime
},s^{\prime }) \sim \Gamma_{c} F(X^{\prime },s), \label{4.19}
\end{equation}
with
\begin{equation}
\Gamma _{c}=  \int_{0}^{\infty }ds^{\prime
}\mathcal{G}_{c}(s^{\prime }). \label{4.20}
\end{equation}

With all these results, the final form for the Fokker-Planck
equation is obtained:
\begin{equation}
\left( \partial _{s}+\gamma \mathbf{V}\cdot \mathbf{\nabla
}_{\mathbf{R} }\right) F(X,s)=\left( \Gamma -\epsilon \Gamma
_{0}\right) \frac{\partial}{\partial {\bi V}}\cdot \left(
\mathbf{V}+\frac{1}{2}\frac{\partial}{\partial {\bi V}} \right)
F(X,s), \label{4.21}
\end{equation}
where
\begin{equation}
\Gamma =\Gamma _{0}+\Gamma _{c}.  \label{4.22}
\end{equation}
is the total friction constant. In this dimensionless form, the
inelasticity of the granular fluid is suppressed and occurs only
through the effective friction constant $\Gamma -\epsilon \Gamma
_{0}$.

\section{Comparison of Granular and Normal Fluids}
\label{s5}

The Fokker-Planck equation plays an important role in
demonstrating explicitly some important general beliefs of
non-equilibrium statistical mechanics. Among these is a two stage
approach to equilibrium. In the present case of a granular fluid,
the role of the equilibrium state is played by the HCS. The first
stage is a fast relaxation of the velocity distribution on the
time scale of a few collisions, followed by a slower transition to
spatial homogeneity and full equilibrium by hydrodynamic
processes. This follows directly from the Fokker-Planck equation,
where the only hydrodynamic process is diffusion.

The recovery of the usual Fokker-Planck equation in dimensionless
variables, equation (\ref{4.21}), means that many of these
qualitative concepts of Brownian motion for normal fluids can be
transferred to granular fluids as well. The general solution
(Green's function) for the Fokker-Planck equation can be
constructed to demonstrate the two stage relaxation. This has been
discussed in some detail for an analysis based on the low density
Boltzmann-Lorentz equation \cite{Dufty02} and the analysis there
applies here as well. The main similarities between normal and
granular fluids are:

\begin{itemize}
\item There is a universal stationary state, Gaussian in the
velocity and spatially uniform, approached at long times for all
physically relevant initial conditions. For initial homogeneous
states, the approach to this stationary state is exponentially
fast in the time scale $s$ determined by the collision number.

\item The spectrum of the Fokker-Planck equation includes an
isolated point representing diffusion. This is the smallest point
in the spectrum, corresponding to the slowest possible excitation.
The next slowest mode decays exponentially fast in $s$ relative to
the diffusion mode. Therefore, for general initial conditions, the
diffusion equation dominates for sufficiently long times. The
existence of diffusion and its dominance applies for all values of
the restitution coefficient $\alpha _{0}$.

\end{itemize}

This result is relevant because the validity of a hydrodynamic
description for granular fluids is not self-evident. Furthermore,
if that validity is granted based on empirical or phenomenological
grounds, it is often assumed to be limited to weak inelasticity.
The Fokker-Planck limit considered here provides an unambiguous
example of the existence of hydrodynamics even for strong
inelasticity, and a description of the initial transient period
leading up to its dominance.

There are also significant differences between normal and granular
fluids that have been suppressed by the use of dimensionless
variables. Returning to the original variables, equation
(\ref{4.21}) takes the form
\begin{equation}
\left( \partial _{t}+\mathbf{V}\cdot \mathbf{\nabla
}_{\mathbf{R}}\right) F(X,t)=\frac{\partial}{\partial {\bi V}}
\cdot \left\{ \Gamma \left[ T(t) \right]
\mathbf{V}+\frac{1}{2}\mathcal{D}\left[ T(t) \right]
\frac{\partial}{\partial {\bi V}} \right\} F(X,t). \label{5.1}
\end{equation}
The first term of the brackets is known as the ``drift'' term and
describes the deterministic dynamics of the particle. The friction
coefficient $\Gamma(t)$ is related with the dimensionless one
$\Gamma^{*}$ appearing in equation (\ref{4.21}) by $\Gamma(t)=
\widetilde{v}(t) \Gamma^{*} / \ell$. The second term represents
diffusion in velocity space due to the finite temperature of the
host fluid and gives rise to all statistical properties of the
dynamics. The velocity diffusion coefficient is given by
\begin{equation}
\mathcal{D}\left[ T(t)\right] =\frac{2 \gamma^{2} T(t)}{m}\left\{
\Gamma \left[ T(t) \right] -\epsilon \Gamma _{0}\left[ T(t)\right]
\right\} .  \label{5.2}
\end{equation}

For Brownian motion in a normal fluid, the corresponding
relationship is $ \mathcal{D}=\left(2 T/m_{0} \right) \Gamma $.
This is known as a fluctuation-dissipation relation (velocity
diffusion representing fluctuations, the friction constant
$\Gamma$ representing dissipation). This relation is violated for
granular fluids in two ways. First, there is the additional term
proportional to $\epsilon $ due to the cooling of the host fluid.
Second, the impurity particle temperature $T_{0}(t)$ differs from
that of the host fluid, as shown in equation (\ref{4.8}) for the
impurity HCS. A related consequence of cooling and the two
different temperatures, is a violation of the Einstein relation
between spatial diffusion and mobility. The latter can be
determined from the Fokker-Planck equation by adding an applied
external force and calculating the response of the average
velocity of the impurity particle. The resulting mobility
coefficient has no simple relationship to the coefficient of
diffusion determined from the mean square displacement of the
particle.

An obvious effect of the presence of a time dependent temperature
in (\ref {5.1}), is a nonlinear change in the relevant time scale.
This follows from the fact that the fluid-fluid collision
frequency sets the physical time scale and this is proportional to
$\sqrt{T(t)}$. Consequently, the dimensionless collision frequency
$s$ depends logarithmically on $t$. This renormalization of the
time is important for identifying the various excitations. For
example, the mean square displacement of the impurity becomes
linear in the collision number time scale $s$, not in $t$.

\section{Discussion}

The Fokker-Planck equation derived here has several qualifications
related to the small parameter $m/m_0$. In fact, there are other
parameters of the problem that can become large such that their
product with the small mass ratio compromises the validity of the
expansion. If the size of the impurity becomes large relative to
the host fluid particle size, the mass densities of the impurity
and fluid particles can be come comparable. In fact, this is a
common experimental condition. In this case, it has been suggested
\cite{Bocquet00} that the Fokker-Planck equation must be modified
to a non-Markovian form, since the time scale separation in the
analysis of the correlated collisions operator $M$ is no longer
justified.

A second possibility is that the ratio of the impurity temperature
to the fluid particle temperature, $T_{0}/T$, becomes large. This
occurs when the cooling rate for the fluid is larger than the
effective impurity-fluid collision rate (violation of the
condition (\ref{4.9}) above). Since the latter decreases with the
mass ratio, this requires that the host fluid particle must be
less inelastic. When this is not the case, the derivation given
here is not valid and the Fokker-Planck description does not
apply. Instead, the impurity particle executes "ballistic" rather
than diffusive motion \cite{SyD01}.

The hard sphere interactions lead to some important differences
from the results for continuous potentials. The collision operator
in equation (\ref{3.4}) has both an instantaneous contribution,
$\Lambda $, and one representing finite time correlated
collisions, $M $. The first is analogous to the
Boltzmann-Enskog-Lorentz collision operator. Accordingly, there
are separate contributions from each of them  to the friction
coefficient, $\Gamma_{0}$ and $\Gamma_{c}$, respectively. This is
puzzling since for continuous potentials the friction coefficient
is given by a single Green-Kubo expression, in terms of the time
integral of the force autocorrelation function. However, it has
been shown recently that this autocorrelation function develops a
singularity for steeply repulsive potentials, leading precisely to
the instantaneous contribution $\Gamma_{0}$ for hard spheres
\cite{DyE04}. The residual non-singular part is given by the
``force autocorrelation function" (\ref{4.17}). This is a general
feature of the hard sphere interaction and is not related to the
inelasticity of the collision.

It is interesting that the operator $M$ describing correlated
many-body collisions plays no role in determining the HCS
distribution for the impurity particle or its cooling rate in the
HCS. On the other hand, it does have an important effect for the
dynamics of deviations from the HCS, through the correlated part
of the friction constant $\Gamma_{c}$. However, the final
temperature of the HCS is independent of this constant.

Within the context described above, the results are exact and
apply for both dense and dilute fluids. Explicit evaluation of the
friction coefficient as a function of the density and restitution
coefficients is a difficult many-body problem. The contribution
$\Gamma_{0}$ is similar to that from the Enskog kinetic theory and
in fact reduces to the latter for elastic collisions. Here, it is
necessary to understand the pair velocity correlations in the HCS
before further simplification is possible. The contribution from
correlated binary collisions $\Gamma_{c}$ is more complicated.
Even at low density and elastic collisions, an infinite sequence
of ``ring'' and ``repeated ring'' recollisions between fluid and
impurity particles must be calculated when the size of the
impurity becomes comparable to the mean free path
\cite{Dorfman80}.

Certainly, the motion of an impurity in a granular fluid provides
conceptual, computational, and experimental challenges for the
second century after Einstein's initial insight.

\ack

The research of J.W.D. was supported in part by Department of
Energy Grant DE-FG02ER54677. The research of J.J.B. was partially
supported by the Ministerio de Ciencia y Tecnolog\'{\i}a (Spain)
through Grant No. BFM2002-00307 (partially financed by FEDER
funds).

\appendix

\section{Formal Kinetic Equation}
\label{ap1} The projection operator method for normal fluids
provides a means to write formally exact equations for observables
of interest in terms of collision kernels. These kernels are then
the appropriate objects to study by means of chosen approximation
schemes \cite{Zw01}. In this Appendix, this method is used to
obtain the exact kinetic equation for the probability distribution
of the impurity particle, defined in equation (\ref{3.3}). The
method has been discussed in detail in many places for normal
fluids and extends directly to the granular fluid when working in
the stationary representation. Only a short summary is given here.

The projection operator over the ``relevant'' part is defined by
\begin{equation}
\mathcal{P} \Phi (\Gamma)=\int dX \rho_{hcs} \left( \Gamma
;X\right) \int d\Gamma ^{\prime }\delta \left(
X\mathbf{-}x_{0}^{\prime }\right)  \Phi \left( \Gamma ^{\prime
}\right), \label{a.1}
\end{equation}
for an arbitrary phase function $\Phi (\Gamma)$. The distribution
function $\rho_{hcs} \left( \Gamma ;X\right)$ is defined in
equation (\ref{3.2}). It is easily verified that $\mathcal{P}$
actually has the projection property $\mathcal{P}^{2}
=\mathcal{P}$. Moreover, it is
\begin{equation}
\int d\Gamma\, \delta(X-x_{0}) \mathcal{P} \rho(\Gamma,s)= \int d
\Gamma\, \delta (X- x_{0}) \rho (\Gamma,s) = F(X,s). \label{a.2}
\end{equation}
Consequently, $\mathcal{P} \rho(\Gamma,t)$ retains the relevant
information in order to evaluate the probability distribution for
the impurity. To obtain an equation for $\mathcal{P}
\rho(\Gamma,t)$, we start from the Liouville equation
(\ref{2.23}), where the binary collision operator for collisions
between the impurity and the fluid particles is
\begin{equation}
\fl \overline{T}_{-}(\gamma;0,i)=\overline{\sigma}^{d-1} \int d
\Omega\, \Theta \left[ \overline{\bi g}_{0i}(\gamma) \cdot
\widehat{\bsigma} \right] |\overline{\bi g}_{0i}(\gamma) \cdot
\widehat{\bsigma}| \left[ \alpha_{0}^{-2} \delta ({\bi
q}_{0i}-\overline{\bsigma}) \overline{b}_{0i}^{-1}-\delta ({\bi
q}_{0i}+\overline{\bsigma}) \right], \label{a.3}
\end{equation}
where
\begin{equation}
\overline{\bi g}_{0i}(\gamma)={\bi v}_{0}- \gamma^{-1} {\bi v}_{1}
\label{a.4}
\end{equation}
and the operator $\overline{b}_{0i}^{-1}$ is defined by
\begin{equation}
\overline{b}_{0i}^{-1} {\bi v}_{0}={\bi
v}_{0}-\frac{(1+\alpha_{0})\Delta}{\alpha_{0}} \left[
\overline{\bi g}_{0i}(\gamma) \cdot \widehat\bsigma \right]
\widehat\bsigma, \label{a.5}
\end{equation}
\begin{equation}
\overline{b}_{0i}^{-1} {\bi v}_{1}={\bi
v}_{1}+\frac{(1+\alpha_{0})(1-\Delta) \gamma}{\alpha_{0}} \left[
\overline{\bi g}_{0i}(\gamma) \cdot \widehat\bsigma \right]
\widehat\bsigma. \label{a.6}
\end{equation}

We operate on both sides of the Liouville equation with the
operator $\mathcal{P}$ and rearrange the result to get
\begin{equation}
\left( \partial
_{s}+\mathcal{P}\overline{\mathcal{L}}\mathcal{P}\right)
\mathcal{P}\rho \left( \Gamma,s\right)
=-\mathcal{P}\overline{\mathcal{L}} \mathcal{Q}\rho \left(
\Gamma;s\right), \label{a.7}
\end{equation}
with $\mathcal{Q} \equiv 1- \mathcal{P}$. Repeat this analysis by
operating on the Liouville equation with $\mathcal{Q}$ and
rearranging in a similar way,
\begin{equation}
\left( \partial
_{s}+\mathcal{Q}\overline{\mathcal{L}}\mathcal{Q}\right)
\mathcal{Q}\rho \left( \Gamma,s\right)
=-\mathcal{Q}\overline{\mathcal{L}} \mathcal{P}\rho \left(
\Gamma,s\right) .  \label{a.8}
\end{equation}
Equations (\ref{a.7}) and (\ref{a.8}) are a coupled pair of
equations for $ \mathcal{P}\rho$ and $\mathcal{Q}\rho$. They are
easily solved with the initial conditions $ \mathcal{P}\rho \left(
\Gamma ;0\right) =\rho \left( \Gamma,0
\right)=\rho_{hcs}(\Gamma;X_{0}) $ and $\mathcal{Q}\rho \left(
\Gamma ;0 \right) =0$, following from equation (\ref{3.2}), with
the result
\begin{equation}
\left( \partial
_{s}+\mathcal{P}\overline{\mathcal{L}}\mathcal{P}\right)
\mathcal{P}\rho \left( \Gamma, s\right) -\int_{0}^{s}ds^{\prime
}\, \mathcal{P
}\overline{\mathcal{L}}\rme^{-(s-s^{\prime})\mathcal{Q}\overline{\mathcal{L}}\mathcal{Q}
}\mathcal{Q}\overline{\mathcal{L}}\mathcal{P} \rho \left( \Gamma,
s^{\prime }\right) =0.  \label{a.9}
\end{equation}
Finally, making some of the projection operators explicit and
using equation (\ref{a.2}), gives the desired kinetic equation
\begin{eqnarray}
\fl \partial _{s}F(X,s)+\int dX^{\prime }B(X,X^{\prime
})F(X^{\prime },s)\nonumber \\
+\int_{0}^{s}ds^{\prime }\int
dX^{\prime }M(X,X^{\prime },s-s^{\prime })F(X^{\prime },s^{\prime
})=0, \label{a.10}
\end{eqnarray}
with
\begin{equation}
B(X,X^{\prime }) \equiv \int d\Gamma \delta \left(
X\mathbf{-}x_{0}\right) \overline{\mathcal{L}}\rho_{hcs} \left(
\Gamma ;X^{\prime }\right),  \label{a.11}
\end{equation}
\begin{equation}
M(X,X^{\prime },s) \equiv -\int d\Gamma \delta \left(
X\mathbf{-}x_{0}\right) \overline{\mathcal{L}} \rme^{-s
\mathcal{Q}\overline{\mathcal{L}}\mathcal{Q}}
\mathcal{Q}\overline{\mathcal{L}}\rho_{hcs} \left( \Gamma
;X^{\prime }\right). \label{a.12}
\end{equation}

The kernel $B(X,X^{\prime })$ represents the instantaneous mean
field effects of the fluid on the particle. This can be made more
transparent by rewriting its expression in the following way:
\begin{equation}
B(X,X^{\prime }) =\int d\Gamma \delta \left(
X\mathbf{-}x_{0}\right) \overline{\mathcal{L}}\delta \left(
X^{\prime }\mathbf{-}x_{0}\right) \rho_{hcs}\left( \Gamma \right)
F_{hcs}^{-1}({\bi V}^{\prime }), \label{a.13}
\end{equation}
where
\begin{equation}
F_{hcs}({\bi V}) \equiv \int d\Gamma\, \delta(X-x_{0})
\rho_{hcs}(\Gamma) \label{a.14}
\end{equation}
is the impurity distribution in the HCS. Then, using equation
(\ref{2.24}), it is found
\begin{equation}
\fl B(X,X^{\prime }) = \gamma \left\{ \left( {\bi V} \cdot
\nabla_{\bi R}+\Lambda \right) \delta \left( X^{\prime
}\mathbf{-}X\right) + \frac{\zeta _{0,hcs}}{2}
\frac{\partial}{\partial {\bi V}} \cdot \left[ {\bi V}\delta
\left( X \mathbf{-}X^{\prime }\right) \right] \right\},
\label{a.15}
\end{equation}
with the linear operator $\Lambda$ defined by
\begin{equation}
\Lambda \delta \left( X^{\prime }\mathbf{-}X\right) =-\int
dx_{1}\overline{T}_{-} (\gamma ;X,x_{1})\delta \left( X^{\prime
}\mathbf{-}X\right) F_{hcs}^{(2)}(X,x_{1})F_{hcs}^{-1}({\bi V}).
\label{a.16}
\end{equation}
Here
\begin{equation}
F_{hcs}^{(2)}(X,x_{1})=N \int dx_{0} \int dx_{2} \cdots \int
dx_{N}\,  \delta(x_{0}-X) \rho _{hcs}\left( \Gamma \right)
\label{a.17}
\end{equation}
is the reduced two particle distribution for the impurity and one
fluid particle, and the operator
$\overline{T}_{-}(\gamma;X,x_{1})$ is still defined by equation
(\ref{a.2}), but replacing $x_{0}$ by $X$. Equation (\ref{3.4})
follows by substituting equation (\ref{a.15}) into equation
(\ref{a.10}). As discussed in the main text, $\Lambda $ is closely
related to the linearized Boltzmann-Enskog-Lorentz operator for a
granular fluid in dimensionless form.

The kernel $M(X,X^{\prime },s)$ is somewhat more complex. If the
complete system is in the HCS, i.e.
$\rho(\Gamma,s)=\rho_{hcs}(\Gamma)$ and, consequently,
$F(X,s)=F_{hcs}({\bi V})$, it is
\begin{equation}
\int dX^{\prime} M(X,X^{\prime},s)F_{hcs}({\bi V}^{\prime})=0,
\label{a.18}
\end{equation}
for all $s$. This is a consequence of the property
\begin{equation}
\int dX^{\prime}\,  \overline{\mathcal{L}}(\Gamma)
\rho_{hcs}(\Gamma,X^{\prime}) F_{hcs}({\bi V}^{\prime})=
\overline{\mathcal{L}} \rho_{hcs}(\Gamma)=0. \label{a.19}
\end{equation}
Simplifications of $M$ occur by writing
\begin{equation}
\overline{\mathcal{L}}=\overline{\mathcal{L}}_{0}+\overline{\mathcal{L}}%
_{f}- \gamma \sum_{i=1}^{N}\overline{T}_{-}(\gamma ;0,i),
\label{a.20}
\end{equation}
where $\overline{\mathcal{L}}_{0}$ and
$\overline{\mathcal{L}}_{f}$ are the generators for the dynamics
of the isolated impurity and the isolated fluid respectively.
Their expressions are directly identified from equation\
(\ref{2.24}). It is easily verified that
\begin{equation}
\mathcal{P}\left(
\overline{\mathcal{L}}_{0}+\overline{\mathcal{L}} _{f}\right)
\mathcal{Q} =0, \label{a.21}
\end{equation}
and equation (\ref{a.12}) thus becomes
\begin{eqnarray}
\fl M(X,X^{\prime },s)=\gamma \int d\Gamma \delta \left( X -
x_{0}\right) \sum_{i=1}^{N}\overline{T}_{-}(\gamma ;0,i) \nonumber
\\
\times \rme^{-s\mathcal{Q} \overline{\mathcal{L}} \mathcal{Q}
}\mathcal{Q}\overline{\mathcal{L}} \rho _{hcs}\left( \Gamma
\right) \delta \left( X^{\prime }\mathbf{-} x_{0}\right)
F_{hcs}^{-1}({\bi V}^{\prime }).  \label{a.22}
\end{eqnarray}
Next, use the identity
\begin{equation}
\fl \overline{\mathcal{L}}\rho _{hcs}\left( \Gamma \right)
\Phi(\Gamma ) =\left[ \overline{\mathcal{L}}\rho _{hcs}\left(
\Gamma \right) \right] \Phi(\Gamma )+\rho _{hcs}\left( \Gamma
\right) \mathcal{L}_{-}\Phi(\Gamma )  = \rho _{hcs}\left( \Gamma
\right) \mathcal{L}_{-}\Phi(\Gamma ), \label{a.23}
\end{equation}
valid for arbitrary $\Phi(\Gamma )$. The new generator
$\mathcal{L}_{-}$ is defined by
\begin{eqnarray}
\fl \mathcal{L}_{-}\Phi(\Gamma) = \left[ \gamma {\bi v}_{0}\cdot
\mathbf{\nabla }_{0}- \gamma \sum_{i=1}^{N}T_{-}(\gamma ;0,i)
+\frac{\zeta _{hcs}}{2} {\bi v}_{0} \cdot \frac{\partial}{\partial
{\bi v}_{0}} \right. \nonumber \\
+ \left. \sum_{i=1}^{N} {\bi
v}_{i}\cdot \mathbf{\nabla
}_{i}-\frac{1}{2}\sum_{i=1}^{N}\sum_{j\neq i}^{N}T_{-}(i,j)
+\frac{\zeta _{hcs}}{2} \sum_{i=1}^{N} {\bi v}_{i} \cdot
\frac{\partial}{\partial {\bi v}_{i}} \right] \Phi
(\Gamma),\label{a.24}
\end{eqnarray}
with the modified binary collision operators
\begin{equation}
T_{-}(i,j)= \sigma^{d-1} \int d\Omega \,\delta
(\mathbf{q}_{ij}-{\bsigma })\Theta (\mathbf{g}_{ij}\cdot
\widehat\bsigma|\mathbf{g} _{ij}\cdot \widehat\bsigma|\left(
b_{ij}^{-1}-1\right), \label{a.25}
\end{equation}
\begin{equation}
T_{-}(\gamma ;0,i)=\overline{\sigma}^{d-1} \int d\Omega \,\delta
(\mathbf{q}_{0i}-\overline{\bsigma} )\Theta
(\overline{\mathbf{g}}_{0i}\cdot
\widehat{\bsigma)}|\overline{\mathbf{g}} _{0i}\cdot
\widehat{\bsigma}|\left( \overline{b}_{0i}^{-1}-1\right).
\label{a.26}
\end{equation}
For systems with elastic collisions, $\mathcal{L}_{-}$ is the
generator for the time reversed dynamics, and it plays a similar
role for the HCS time correlation functions of granular systems.
The detailed proof of equation (\ref{a.24}) will be given
elsewhere. With this identity, equation\
 (\ref{a.22}) can be
rewritten as
\begin{eqnarray}
\fl M(X,X^{\prime },s) = -\gamma^{2} \int d\Gamma \delta \left(
X\mathbf{-}x_{0}\right) \sum_{i=1}^{N}\overline{T}_{-}(\gamma
;0,i)\rme^{-s \mathcal{Q}\overline{\mathcal{L}} \mathcal{Q}
}\mathcal{Q}\rho _{hcs}\left( \Gamma
\right)   \nonumber \\
\times \sum_{j=1}^{N}T_{-}(\gamma ;0,j)\delta \left( X^{\prime
}\mathbf{-} x_{0}\right) F_{hcs}^{-1}({\bi V}^{\prime }).
\label{a.27}
\end{eqnarray}

\section{Mass Ratio Expansion}
\label{ap2} The expressions for $\Lambda $ and $M$ are given in
Eqs.\ (\ref{a.16}) and (\ref{a.27}), respectively. With the
appropriate choice of $H(x_{0},x_{1})$) both have the form
\begin{equation}
Z \equiv \int dx_{0}\, \delta (X-x_{0})\int dx_{1}\,
\overline{T}_{-}(\gamma ;0,1)H(x_{0},x_{1}), \label{b.1}
\end{equation}
that using the expression of the binary collision operator can be
transformed into
\begin{eqnarray}
\fl Z  =  \int dx_{0}\int dx_{1}\delta \left( X-x_{0}\right)
\overline{\sigma}^{d-1} \int d\Omega \,\Theta
(\overline{\mathbf{g}} _{01}\cdot \widehat{\mbox{\boldmath
$\sigma$}})|\overline{\mathbf{g}}_{01}\cdot
\widehat{\mbox{\boldmath $\sigma$}}|  \nonumber \\
\times  \left[ \alpha _{0}^{-2}\delta (\mathbf{q}_{01}-\overline{
\mbox{\boldmath $\sigma$}})b_{01}^{-1}-\delta (\mathbf{q}_{01}+
\overline{\mbox{\boldmath $\sigma$}})\right] H(x_{0},x_{1})
\nonumber \\
\lo= \int dx_{0} \int dx_{1}H(x_{0},x_{1})\delta \left(
\mathbf{R-r}_{0}\right) \overline{\sigma}^{d-1} \int d\Omega
\,\Theta (\overline{\mathbf{ g}}_{01}\cdot
\widehat{\mbox{\boldmath
$\sigma$}})|\overline{\mathbf{g}}_{01}\cdot
\widehat{\mbox{\boldmath $\sigma$}}|\delta
(\mathbf{q}_{01}+\overline{\mbox{\boldmath $\sigma$}}) \nonumber
\\
\times \left[ \delta \left( {\bi V}-\overline{b}_{01} {\bi
v}_{0}\right) -\delta \left( {\bi V}-{\bi v}_{0}\right) \right],
\label{b.2}
\end{eqnarray}
where $\overline{b}_{01}{\bi v}_{0}$ is obtained from equation
(\ref{a.5}),
\begin{equation}
\overline{b}_{01}{\bi v}_{0}= {\bi v}_{0}-\left( 1+\alpha _{0}
\right) \Delta (\overline{\mathbf{g}}_{01} \cdot \widehat\bsigma
)\widehat\bsigma. \label{b.3}
\end{equation}
In the limit of a very massive impurity, it is $\Delta \gamma
^{-1}<<1$, and $\delta ({\bi V}-\overline{b}_{01} {\bi v}_{0} )$
in equation (\ref{b.2}) can be expanded to second order yielding
\begin{eqnarray}
\fl Z \sim  \int dx_{0}\int dx_{1}H(x_{0},x_{1})\delta \left(
\mathbf{R-r} _{0}\right) \nonumber \\
\times  \left[ - \mathbf{A} (x_{0},x_{1}) \cdot
\frac{\partial}{\partial {\bi v}_{0}} \delta \left( {\bi V}-{\bi
v}_{0}\right) +\frac{1}{2} \textsf{B}(x_{0},x_{1})
:\frac{\partial}{\partial {\bi v}_{0}}\frac{\partial}{\partial
{\bi v}_{0}} \delta \left( {\bi V}-{\bi v}_{0} \right) \right]
\nonumber \\
\lo= \int dx_{0}\delta (X-x_{0}) \nonumber \\
\times \int dx_{1}\left[ \frac{\partial}{\partial {\bi v}_{0}}
\cdot \mathbf{A}(x_{0},x_{1})+\frac{1}{2} \frac{\partial}{\partial
{\bi v}_{0}} \frac{\partial}{\partial {\bi v}_{0}} :
\textsf{B}(x_{0},x_{1})\right] H(x_{0},x_{1}), \label{b.4}
\end{eqnarray}
with the definitions
\begin{equation}
\fl A_{i}(x_{0},x_{1})=\left( 1+\alpha _{0}\right) \Delta\,
\overline{\sigma}^{d-1}\int d\Omega \, \Theta \left[
\overline{\mathbf{g}}_{01}(\gamma) \cdot \widehat{\mbox{\boldmath
$\sigma$}} \right] \delta ( {\bi q}_{01}+ \overline{\bsigma} )
[\overline{\bi g}_{01}(\gamma) \cdot \widehat\bsigma]^{2}
\sigma_{i}, \label{b.5}
\end{equation}
\begin{equation}
\fl B_{ij}(x_{0},x_{1})=\left[(1+\alpha_{0})\Delta \right]^{2}
\overline{\sigma}^{d-1} \int d\Omega \,\Theta
[\overline{\mathbf{g}}_{01} (\gamma) \cdot \widehat{
\mbox{\boldmath $\sigma$}}] \delta
(\mathbf{q}_{01}+\overline{\mbox{\boldmath
$\sigma$}})|\overline{\bi g}_{01}(\gamma) \cdot
\widehat\bsigma|^{3} \widehat{{\sigma }}_{i}\widehat{ {\sigma
}}_{j}\, . \label{b.6}
\end{equation}
Comparison of Eqs.\ (\ref{b.1}) and (\ref{b.4}) gives the relation
\begin{eqnarray}
\fl \int dx_{1}\overline{T}_{-}(\gamma ;0,1)H(x_{0},x_{1})  \sim
\frac{\partial }{\partial {\bi v}_{0}} \cdot \left[ \int dx_{1}
{\bi A}(x_{0},x_{1})H(x_{0},x_{1})  \right. \nonumber \\
  \left. + \frac{1}{2}\frac{\partial }{\partial {\bi v}} \cdot
\int dx_{1} \textsf{B}(x_{0},x_{1})H(x_{0},x_{1})\right] .
\label{b.7}
\end{eqnarray}
It should be noted that this is not the complete expansion in the
mass ratio, since there remains a dependence on $\gamma$.

\subsection{Evaluation of $\Lambda $}

For the particular case of $\Lambda $, equation (\ref{b.7}) gives
\begin{eqnarray}
\fl \Lambda F(X,s) =-\int dx_{1}\overline{T}_{-}(\gamma
;X,x_{1})F_{hcs}^{(2)}(X,x_{1})F_{hcs}^{-1}({\bi V})F(X,s)  \nonumber \\
\lo{\sim} -\frac{\partial }{\partial {\bi V}} \cdot \left[
\overline{\bi A}(X)+\frac{1}{2} \frac{\partial }{\partial {\bi
V}}\cdot \overline{\textsf B}(X)\right] F(X,s), \label{b.8}
\end{eqnarray}
with
\begin{eqnarray}
\fl \overline{A}_{i}(X) = \int dx_{0}\, \delta (x_{0}-X) \int
dx_{1}
A_{i}(x_{0},x_{1})F_{hcs}^{(2)}(X,x_{1})F_{hcs}^{-1}({\bi V})  \nonumber \\
\lo= \left( 1+\alpha _{0}\right) \Delta \left( \frac{\overline{
\sigma }}{\sigma }\right) ^{d-1}  \int dx_{0} \delta (x_{0}-X)
\int d {\bi v}_{1}\varphi_{hcs}({\bi v}_{1})\int d\Omega \,\Theta
(\overline{\mathbf{g}}_{01}\cdot \widehat{\bsigma})
\nonumber \\
\times (\overline{\bi g}_{01} \cdot \widehat{{\bsigma }
})^{2}g^{(2)} (\overline{\bsigma},{\bi V},{\bi v}_{1})
\widehat{{\sigma }}_{i}, \label{b.9}
\end{eqnarray}
\begin{eqnarray}
\fl \overline{B}_{ij}(X) = \int dx_{0}\, \delta (x_{0}-X) \int
dx_{1} B_{ij}(x_{0},x_{1})F_{hcs}^{(2)}(X,x_{1})F_{hcs}^{-1}({\bi V})  \nonumber \\
\lo= [ \left( 1+\alpha _{0} \right) \Delta ]^{2} \left(
\frac{\overline{\sigma }}{\sigma }\right) ^{d-1} \int dx_{0}\,
\delta (x_{0}-X) \int d {\bi v} _{1} \varphi_{hcs}({\bi
v}_{1})\int d\Omega\, \Theta (\overline{\mathbf{g}}_{01} \cdot
\widehat{\bsigma}) \nonumber \\
\times (\overline{\mathbf{g}}_{01} \cdot
\widehat{\bsigma})^{3}g^{(2)}(\overline{\bsigma}, {\bi
 V}, {\bi v} _{1})
\widehat{{ \sigma }}_{i}\widehat{{\sigma }}_{j}. \label{b.10}
\end{eqnarray}
In the above expressions, we have introduced $\varphi_{hcs}({\bi
v}_{1}) \equiv n^{-1} f_{hcs}({\bi v}_{1})$, where $f_{hcs}({\bi
v}_{1})$ is the one-particle distribution of the fluid in the HCS
and the correlation function $g^{(2)}(\overline{\bsigma},{\bi V},
{\bi v}_{1})$ between the impurity and the fluid particles,
defined through
\begin{equation}
F_{hcs}^{(2)}(\overline{\bsigma}, {\bi V},{\bi
v}_{1})=f_{hcs}({\bi v}_{1})F_{hcs}({\bi
 V})g^{(2)}(\overline{\bsigma}, {\bi
 V},{\bi v}_{1} ). \label{b.11}
\end{equation}
Because of the spatial homogeneity and isotropy of the HCS, the
correlation function must be a function of the scalars
$\overline{\sigma},V,v_{1}, \widehat{\bi
 V} \cdot \widehat\bsigma, \widehat{{\bi v}} _{1} \cdot \widehat\bsigma,$ and
$\widehat{\bi
 V}\cdot \widehat{{\bi v}}_{1}$, where the hats
indicate unit vectors. Furthermore, the characteristic impurity
velocity is a factor $\gamma $ smaller than the characteristic
fluid velocity. Therefore, the correlation function may be
expanded to first order as
\begin{equation}
g^{(2)}(\overline{\bsigma},{\bi V},{\bi v}_{1}) \sim
g^{(2)}_{0}(\overline{\sigma} ,v_{1}, \widehat{\bi v}_{1} \cdot
\widehat{\bsigma})+\gamma {\bi V} \cdot {\bi g}^{(2)}_{1}(\sigma
,v_{1}, \widehat{\bi v}_{1} \cdot \widehat{\bsigma}). \label{b.12}
\end{equation}
Now equation (\ref{b.9}) can be written
\begin{equation}
\overline{\bi A}(X)=\gamma^{-1} \Gamma _{0} {\bi V}, \quad \quad
\Gamma _{0}=\Gamma _{0}^{(0)}+\Gamma _{0}^{(1)}, \label{b.13}
\end{equation}
where
\begin{eqnarray}
\fl \Gamma _{0}^{(0)} \sim  \left( 1+\alpha _{0}\right) \Delta
\left( \frac{\overline{\sigma}}{\sigma} \right)^{d-1} \int d{\bi
v}_{1}\varphi_{hcs}({\bi v}_{1})\int d\Omega \,\Theta (-{\bi
v}_{1} \cdot \widehat\bsigma ) g^{(2)}_{0}(\overline{\sigma}
,v_{1}, \widehat{\bi v}_{1} \cdot \widehat\bsigma)(\widehat{\bi V}
\cdot
\widehat{\bsigma })  \nonumber \\
\times \left[ \gamma^{-1} V^{-1}({\bi v}_{1} \cdot
\widehat{\bsigma })^{2}- 2( {\bi v}_{1} \cdot \widehat{\bi
 \sigma
})(\widehat{\bi V} \cdot \widehat{\bsigma })\right]. \label{b.14}
\end{eqnarray}
The term of order $\gamma ^{-1}$ is of odd parity in ${\bi v}_{1}$
(change ${\bi v}_{1}\rightarrow -{\bi v}_{1},\widehat\bsigma
\rightarrow -\widehat{\bsigma }$) and, therefore, it vanishes when
integrated over the isotropic distribution $\varphi_{hcs}({\bi
v}_{1})$. This leaves the dominant term as
\begin{eqnarray}
\fl \Gamma_{0}^{(0)}  =  -2 \left( 1+\alpha _{0}\right) \Delta
\left( \frac{\overline{\sigma }}{\sigma }\right) ^{d-1}\int d{\bi
v} _{1}\varphi_{hcs}({\bi v}_{1})\int d\Omega\, \Theta (-{\bi
v}_{1}\cdot \widehat\bsigma) \nonumber
\\
\times ({\bf v}_{1} \cdot \widehat\bsigma)( \widehat{\bi V} \cdot
\widehat\bsigma)^{2} g^{(2)}_{0}(\overline{\sigma}
,v_{1},\widehat{\bi v}_{1} \cdot \widehat\bsigma). \label{b.15}
\end{eqnarray}
Similarly, the contribution to $\Gamma $ from
$\mathbf{g}^{(2)}_{1}$ to leading order is
\begin{eqnarray}
\fl \Gamma _{0}^{(1)} = \left( 1+\alpha _{0}\right) \Delta \left(
\frac{ \overline{\sigma }}{\sigma }\right) ^{d-1} \int d{\bi
v}_{1} \varphi_{hcs}({\bi v}_{1}) \nonumber \\
\times \int d\Omega \,\Theta (-{\bi
 v}_{1}\cdot \widehat{ \mbox{\boldmath $\sigma$}})({\bi v}_{1}
\cdot \widehat\bsigma )^{2}( \widehat{\bi V} \cdot
\widehat\bsigma)\widehat{\bi V} \mathbf{\cdot
g}^{(2)}_{1}(\overline{\sigma} ,v_{1},\widehat{\bi v}_{1} \cdot
\widehat\bsigma). \label{b.16}
\end{eqnarray}
Combining these gives
\begin{equation}
\Gamma _{0}= \frac{\left( 1+\alpha _{0}\right) \Delta }{2}\left(
\frac{ \overline{\sigma }}{\sigma }\right) ^{d-1}C_{1}\rightarrow
\frac{m}{m_{0}} \frac{\left( 1+\alpha _{0}\right) }{2}\left(
\frac{\overline{\sigma }}{ \sigma }\right) ^{d-1}C_{1},
\label{b.17}
\end{equation}
where $C_{1}$ is a pure number,
\begin{eqnarray}
\fl C_{1} = 4\int d {\bi v}_{1}\varphi_{hcs}({\bi v}_{1}) \int
d\Omega \,\Theta (- {\bi v }_{1} \cdot \widehat{ \bsigma})({\bi
v}_{1} \cdot \widehat\bsigma)(\widehat{\bi V}
\cdot \widehat\bsigma)  \nonumber \\
\times \left[ -(\widehat{\bi V} \cdot \widehat\bsigma
)g^{(2)}_{0}(\overline{\sigma} ,v_{1}, \widehat{\bi v}_{1} \cdot
\widehat\bsigma)+( {\bi v}_{1} \cdot \widehat\bsigma)\widehat{\bi
V}\mathbf{\cdot g} ^{(2)}_{1}(\overline{\sigma} ,v_{1},
\widehat{\bi v} _{1} \cdot \widehat\bsigma)\right]. \label{b.18}
\end{eqnarray}

The analysis of $B_{ij}(X)$ follows in the same way, with the
dominant contribution being:
\begin{equation}
B_{ij}(X)= \gamma^{-1} G \Gamma_{0} \delta _{ij},  \label{b.19}
\end{equation}
where
\begin{equation}
G=\frac{\gamma^{-2}\left( 1+\alpha _{0}\right)  \Delta}{2}\, C_{2}
\label{b.20}
\end{equation}
and $C_{2}$ is another pure number,
\begin{equation}
C_{2}=\frac{4}{C_{1}d}\int d {\bi v}_{1}\varphi_{hcs}({\bi
v}_{1})\int d\Omega \,\Theta ({\bi v}_{1}\cdot
\widehat{\mbox{\boldmath $\sigma$}})( {\bi v}_{1} \cdot
\widehat\bsigma)^{3}g^{(2)}_{0}(\overline{\sigma}
,v_{1},\widehat{\bi v}_{1} \cdot \widehat\bsigma), \label{b.21}
\end{equation}
so that $G$ is independent of the mass ratio in the Brownian limit
we are considering.

\subsection{Evaluation of $M$}

To evaluate $M(X,X^{\prime },s)$ in the small mass ratio limit,
first equation (\ref{b.7}) is used into equation (\ref{a.27}) to
get
\begin{eqnarray}
\fl M(X,X^{\prime },s) \sim - \gamma^{2} \int d\Gamma \delta
\left( X\mathbf{-} x_{0}\right) \frac{\partial}{\partial {\bi
v}_{0}} \cdot \sum_{i=1}^{N} {\bi A}
(x_{0},x_{i})\rme^{-s\mathcal{Q}\overline{\mathcal{L}}\mathcal{Q}
}\mathcal{Q}\rho _{hcs}\left( \Gamma \right) \nonumber
\\
\times \sum_{j=1}^{N}T_{-}(\gamma ;0,j)\delta \left( X^{\prime
}\mathbf{-}
x_{0}\right) F_{hcs}^{-1}({\bi V}^{\prime })  \nonumber \\
\lo= - \gamma \frac{\partial}{\partial {\bi V}} \cdot \int d\Gamma
\delta \left( X\mathbf{-} x_{0}\right) {\bi F}_{+}(\Gamma)
\rme^{-s
\mathcal{Q}\overline{\mathcal{L}}\mathcal{Q}}\mathcal{Q}\rho
_{hcs}\left( \Gamma \right)
\nonumber \\
\times \sum_{j=1}^{N}T_{-}(\gamma ;0,j)\delta \left( X^{\prime
}\mathbf{-} x_{0}\right) F_{hcs}^{-1}({\bi V}^{\prime }),
\label{b.22}
\end{eqnarray}
where we have defined
\begin{eqnarray}
\fl {\bi F}_{+}(\Gamma)= \gamma \sum_{i=1}^{N} {\bi
A}(x_{0},x_{i})= \gamma \left( 1+\alpha _{0}\right) \Delta
\overline{\sigma}^{d-1} \sum_{i=1}^{N}\int d\Omega \,\Theta
(\overline{\mathbf{g}}_{0i}\cdot \widehat{\mbox{\boldmath
$\sigma$}})\delta (\mathbf{q}_{0i}+\overline{ \mbox{\boldmath
$\sigma$}})( \overline{\mathbf{g}}_{0i} \cdot \widehat\bsigma)^{2}
\widehat{\bsigma} \nonumber \\
\lo{\sim}  \gamma^{-1}  \left( 1+\alpha _{0}\right) \Delta
\overline{\sigma}^{d-1} \sum_{i=1}^{N}\int d\Omega \,\Theta (-{\bi
v}_{i} \cdot \widehat{\mbox{\boldmath $\sigma$}})\delta
(\mathbf{q}_{0i}+\overline{ \mbox{\boldmath $\sigma$}})( {\bi
v}_{i} \cdot \widehat\bsigma)^{2}\widehat\bsigma. \label{b.23}
\end{eqnarray}
Upon writing equation (\ref{b.22}), we have neglected the term
proportional to $\textsf{B}$ in equation (\ref{b.7}), because it
is of the order $\gamma^{-3} \Delta^{2}$, while ${\bi A}$ is of
the order $\gamma^{-2} \Delta$. In the case of $\Lambda$, the
situation was different because the leading contribution from
${\bi A}$ happened to identically vanish.

The action of the operator $T_{-}(\gamma ;0,j)$ on $\delta \left(
X^{\prime } \mathbf{-}x_{0}\right) $ can be calculated to leading
order for a similar analysis to the one carried out at the
beginning of this Section, obtaining
\begin{eqnarray}
\fl T_{-}(\gamma ;0,j)\delta \left( X^{\prime
}\mathbf{-}x_{0}\right) = \overline{\sigma }^{d-1} \int d\Omega\,
\delta ( \mathbf{q}_{0j}-\overline{\bsigma})\Theta
(\overline{\mathbf{g}}_{0j}\cdot
\widehat\bsigma)|\overline{\mathbf{g}}_{0j}\cdot \widehat\bsigma|
\nonumber \\
\times \left[ \delta \left( X^{\prime
}-\overline{b}_{0j}^{-1}x_{0}\right)
-\delta \left( X^{\prime }- x_{0}\right) \right]   \nonumber \\
\lo{\sim} - \frac{(1+\alpha_{0})\Delta}{\alpha_{0}}\,  \overline{
\sigma}^{d-1} \int d\Omega \,\delta (\mathbf{q}_{0j}-
\overline{\bsigma})\Theta (\overline{\mathbf{g}}_{0j}\cdot
\widehat{\bsigma})(\overline{\mathbf{g}}_{0j} \cdot
\widehat{\bsigma})^{2} \nonumber \\
\times\,  \widehat{\bsigma} \cdot \frac{\partial}{\partial {\bi
V}^{\prime}} \delta (X^{\prime}-x_{0}). \label{b.24}
\end{eqnarray}
Then,
\begin{eqnarray}
\fl M(X,X^{\prime },s)  \sim  - \frac{\partial}{\partial {\bi V}}
\cdot \int d\Gamma \delta \left( X-x_{0}\right) {\bi
F}_{+}(\Gamma) \rme^{-s \mathcal{Q}\overline{
\mathcal{L}}\mathcal{Q} }\mathcal{Q}\rho _{hcs}\left( \Gamma
\right) {\bi F}_{-}(\Gamma) \nonumber \\
\times F_{hcs}^{-1}({\bi V}^{\prime})\cdot
\frac{\partial}{\partial {\bi V^{\prime}}} \delta \left( X^{\prime
}- x_{0}\right), \label{b.25}
\end{eqnarray}
with
\begin{eqnarray}
\fl {\bi F}_{-}(\Gamma) = \frac{\gamma (1+\alpha_{0})
\Delta}{\alpha_{0}}\, \overline{\sigma }^{d-1} \sum_{i=1}^{N}\int
d\Omega \,\Theta (-\overline{\mathbf{g}}_{0i}\cdot
\widehat\bsigma)\delta ( \mathbf{q}_{0i}+\overline{\mbox{\boldmath
$\sigma$}})( \overline{\mathbf{g}}_{0i} \cdot \widehat\bsigma)^{2}\widehat{\bsigma}
\nonumber \\
\lo{\sim}  \frac{\gamma^{-1} (1+\alpha_{0}) \Delta}{\alpha_{0}}\,
\overline{\sigma }^{d-1} \sum_{i=1}^{N}\int d\Omega \,\Theta ({\bi
v}_{i} \cdot \widehat\bsigma)\delta (
\mathbf{q}_{0i}+\overline{\mbox{\boldmath $\sigma$}})( {\bi v}_{i}
\cdot \widehat\bsigma)^{2}\widehat\bsigma . \label{b.26}
\end{eqnarray}
Since both ${\bi F}_{+}(\Gamma)$ and ${\bi F}_{-}(\Gamma)$ are
proportional to $ \gamma ^{-1} \Delta$, $M(X,X^{\prime },s)$ is of
order $\left(\gamma ^{-1} \Delta \right) ^{2}\sim m/m_{0}$. This
is the same order in the mass ratio as the leading contribution to
$\Lambda $. Therefore, all additional corrections on the mass
ratio in equation (\ref{b.25})  should be neglected by
consistency. In particular this means we can substitute
\begin{equation}
\rme^{- s \mathcal{Q}\overline{\mathcal{L}}\mathcal{Q}}
\mathcal{Q}=\rme^{-s \mathcal{Q}\overline{\mathcal{L}}
}\mathcal{Q} \sim \rme^{-s \mathcal{Q}
\overline{\mathcal{L}}_{f}^{\prime}}\mathcal{Q}= \rme^{-s
\overline{\mathcal{L}}_{f}^{\prime}} \mathcal{Q} \label{b.27}
\end{equation}
where $\overline{\mathcal{L}}_{f}^{\prime}$ is the generator for
the fluid dynamics, including the interactions of the particles
with the impurity at rest. More explicitly,
\begin{equation}
\overline{\mathcal{L}}_{f}^{\prime}=\overline{\mathcal{L}}_{f}-\sum_{i=1}^{N}
\overline{T}^{\prime}_{-}(0,i), \label{b.28}
\end{equation}
where $\overline{\mathcal{L}}_{f}$ is the generator for the
isolated fluid and
\begin{equation}
\fl \overline{T}_{-}^{\prime}(0,i)=\overline{\sigma}^{d-1} \int
d\Omega\, \Theta (-{\bi v}_{1} \cdot \widehat\bsigma) | {\bi
v}_{1} \cdot \widehat\bsigma| \left[ \alpha_{0}^{-2} \delta ( {\bi
q}_{0i}-\overline{\bsigma})b^{\prime -1}_{i}+\delta ( {\bi
q}_{0i}+\overline{\bsigma}) \right], \label{b.29}
\end{equation}
\begin{equation}
b^{\prime -1}_{i} {\bi v}_{i}={\bi
v}_{i}-\frac{1+\alpha_{0}}{\alpha_{0}}\, ({\bi v}_{i} \cdot
\widehat\bsigma) \widehat\bsigma. \label{b.30}
\end{equation}
Since $\overline{\mathcal{L}}^{\prime}$ does not contain dynamics
for the impurity, equation (\ref{b.25}) can be transformed into
\begin{eqnarray}
\fl M(X,X^{\prime},s)= -\frac{\partial}{\partial {\bi V}} \cdot
\int d\Gamma\, {\bi F}_{+}(\Gamma)
\rme^{-s\overline{\mathcal{L}}^{\prime}_{f}}
\rho_{hcs}(\Gamma){\bi F}_{-} (\Gamma) \nonumber \\
\times \delta (X-x_{0}) F^{-1}_{hcs} ({\bi V}^{\prime}) \cdot
\frac{\partial}{\partial {\bi V}^{\prime}} \delta (X-X^{\prime}),
\label{b.31}
\end{eqnarray}
where the relation $ \mathcal{P} \rho_{hcs}(\gamma) {\bi
F}_{-}(\Gamma) \delta (X^{\prime}-x_{0})=0$ has been used. Then,
\begin{equation}
\fl \int dX^{\prime} M(X,X^{\prime},s-s^{\prime})
F(X^{\prime},s^{\prime})= \frac{\partial}{\partial {\bi V}} \cdot
\textsf{G}_{c}(s-s^{\prime}) \cdot \left(  {\bi V}+\frac{1}{2}\,
\frac{\partial}{\partial {\bi V}} \right) F(X,s^{\prime}),
\label{b.32}
\end{equation}
with
\begin{equation}
\textsf{G}_{c}(s)= 2 \int d\Gamma\, {\bi F}_{+}(\Gamma) \rme^{-s
\overline{\mathcal{L}}^{\prime}_{f}} \rho_{hcs}(\Gamma) {\bi
F}_{-}(\Gamma) \delta (X-x_{0}) F^{-1}_{hcs} ({\bi V}).
\label{b.33}
\end{equation}
Because of the homogeneity and isotropy of the HCS, it is
\begin{equation}
\textsf{G}_{c}(s)= \mathcal{G}_{c}(s) \textsf{I}, \label{b.34}
\end{equation}
where $\textsf{I}$ is the unit tensor, and
\begin{equation}
\mathcal{G}_{c}(s)=\frac{2}{d} \int d \Gamma_{f} \left[ {\bi
F}_{+}(\Gamma) \cdot \rme^{-s \overline{\mathcal{L}}^{\prime}_{f}}
\rho_{hcs}(\Gamma_{f}|X) {\bi F}_{-}(\Gamma)\right]_{x_{0}=X}.
\label{b.35}
\end{equation}
Here $\Gamma_{f}$ is the phase space associated to the fluid
particles and
\begin{equation}
\rho(\Gamma_{f}|X) = \left[\rho_{hcs}(\Gamma)\right]_{x_{0}=X}
F_{hcs}^{-1}({\bi V}) \label{b.36}
\end{equation}
is the conditional HCS probability distribution of the fluid
particles given that the phase space coordinates of the impurity
are $X \equiv \{ {\bi R},{\bi V} \}$. To leading order in the mass
ratio, we can neglect the velocity correlations between the fluid
particles and the impurity, as already discussed in the evaluation
of $\Lambda$. Then, $\mathcal{G}_{c}(s)$ becomes independent of
$X$. Moreover, in the limit $\alpha \rightarrow 1$,
$\rho_{hcs}(\Gamma_{f}|X)$ can be substituted by its equilibrium
form. In this way, equation (\ref{b.32}) becomes
\begin{equation}
\fl \int dX^{\prime} M(X,X^{\prime},s-s^{\prime})
F(X^{\prime},s^{\prime})= \frac{\partial}{\partial {\bi V}} \cdot
\left(  {\bi V}+ \frac{1}{2}\, \frac{\partial}{\partial {\bi V}}
\right) \mathcal{G}_{c}(s-s^{\prime}) F(X,s^{\prime}).
\label{b.37}
\end{equation}
This is the expression (\ref{4.16}) quoted in the main text.

\section{Evaluation of the Cooling Rate of the fluid}
\label{ap3} The dimensionless cooling rate for fluid in the HCS
defined by Eqs.\ (\ref{2.10}) and (\ref{2.16}) can be reduced to
an integral over the appropriate reduced distribution functions.
The details are given in the Appendix A of reference
\cite{Dufty02}, with the result
\begin{eqnarray}
\fl \zeta_{hcs}=\frac{1}{2dn }\left( 1-\alpha ^{2}\right)
\sigma^{d-1} \int d {\bi v} _{1}\int d{\bi v}_{2}\int d\Omega\,
f_{hcs}^{(2)}({\bi v}_{1},{\bi v}_{2},
\mathbf{q}_{12}=\mbox{\boldmath $\sigma$}) \nonumber \\
\times \Theta (-\mathbf{g}_{12}\cdot \widehat{\mbox{\boldmath
$\sigma$}})|\mathbf{g}_{12}\cdot \widehat{ \mbox{\boldmath
$\sigma$}}|^{3},  \label{c.1}
\end{eqnarray}
where the fluid two-particle distribution function,
\begin{equation}
f_{hcs}^{(2)}(x_{1},x_{2})=N(N-1)\int dx_{0} \int dx_{3} \cdots
\int dx_{N} \rho _{hcs}\left( \Gamma \right), \label{c.2}
\end{equation}
has been introduced. Due to the condition $\epsilon <1$, equation
(\ref{4.9}), the restitution coefficient $\alpha$ for the
fluid-fluid particle collisions must approach unit as the mass
ratio goes to zero. Hence,
\begin{equation}
\fl f^{\ast (2)}\left( {\bi v}_{1},{\bi v}_{2},\mathbf{q}_{12}=
\mbox{\boldmath $\sigma$}\right) \rightarrow \chi (\sigma
)\varphi_{hcs} ({\bi v}_{1})\varphi_{hcs} ({\bi v}_{2}), \quad
\quad \varphi_{hcs} (\mathbf{v}_{i})=\frac{1 }{\pi
^{d/2}}\rme^{-v_{i}^{2}}, \label{c3}
\end{equation}
where $\chi (\sigma )$ is the equilibrium pair correlation
function at contact. Then, the integrals in equation (\ref{c.1})
can be performed, and the leading contribution to the cooling rate
becomes
\begin{equation}
\zeta _{hcs} \sim \frac{4 \pi^{(d-1)/2}}{\sqrt{2} \Gamma(d/2) d}
(1-\alpha) \chi_{0}(\overline{\sigma}).
\end{equation}

\end{document}